\documentclass{iopart}

\usepackage{graphicx,epstopdf}
\usepackage{psfrag}

\renewcommand{\u}{\underline}
\renewcommand{\b}{\mathbf}

\renewcommand{\i}{\includegraphics}

\begin{document}

\title{Inclusion of intersite spatial correlations in the alloy analogy approach to the half-filled ionic Hubbard model}

\author{D~A~Rowlands and Yu-Zhong~Zhang$^{*}$}
\address{Shanghai Key Laboratory of Special Artificial Microstructure Materials and Technology,
School of Physics, Science and Engineering, Tongji University, Shanghai 200092, P.R.~China}
\ead{$^{*}$yzzhang@tongji.edu.cn}

\begin{abstract}
Using the nonlocal coherent-potential approximation we study the effect of intersite spatial correlations on the transition from band insulator to metal as well as from metal to Mott insulator in the ``alloy analogy'' approach to the paramagnetic solution of the half-filled ionic Hubbard model. We find that intersite spatial correlations enhance the metallic phase.
\end{abstract}


%--------------------------------------------------------------------------------------------------------------------------------------------
\section{Introduction}

The Hubbard model of interacting electrons and their associated correlation effects in narrow $d$ and $f$ energy bands~\cite{Hubbard1} has attracted a huge amount of theoretical and computational effort over a number of decades. However, an early theory due to Hubbard in his third paper on the electron correlation problem remains of significant importance since it was the first to predict a Mott metal-insulator transition (MIT) at a finite critical value of the on-site interaction $U$~\cite{Hubbard2}. By making an analogy with disordered alloys (the ``alloy analogy'' (AA)), Hubbard used a Green's function decoupling method to derive a ``scattering correction'' to his earlier Hubbard I approximation~\cite{Hubbard1}, and his solution has been shown~\cite{Velicky1} to be equivalent to the coherent potential approximation (CPA) which was not actually introduced for disordered systems until a few years later~\cite{Soven1}. Hubbard also added a ``resonance broadening'' correction to the AA and this is referred to as the full Hubbard III approximation~\cite{Hubbard2}.

Hubbard's AA improves upon the Hartree-Fock approximation by considering classical or static fluctuations in the potential that an electron sees~\cite{Gyorffy1}, and remains useful as a computationally simple theory capable of giving a valid approximate description of the MIT. Nevertheless, one drawback of Hubbard's work is that the scattering correction obtained from his solution to the AA (i.e.~the CPA) is a local approximation, meaning that intersite or nonlocal spatial correlations in the required alloy configurational average are neglected. Previous work on addressing this issue has been carried out many years ago by Mors~\etal\cite{Mors1} who incorporated a crude two-site spatial correlation in order to study nonlocal effects on the MIT, and more recently by Luo~\etal\cite{Luo2}  based on a combination of the full Hubbard III approximation with the spectral density approach~\cite{Geipel1,Nolting1} and a nonlocal self-energy ansatz. In the present paper we solve the AA using the nonlocal CPA~\cite{Jarrell1,Moradian1,Rowlands3,Rowlands5} which is now an established mean-field approach for describing nonlocal physics and short-range order~\cite{Rowlands1}. Like the CPA it requires the solution of a self-consistent impurity problem but with the added capability of including intersite spatial correlations to arbitrary length scale depending on the size of the impurity cluster used in the calculation.

We study the interesting case of the ionic Hubbard model~\cite{Hubbard3} (IHM) which adds a staggered ionic potential to the conventional Hubbard model on a bipartite lattice i.e.~two sublattices with different on-site energies. This enables us to examine nonlocal effects on the expected transition from band insulator to metal as well as the MIT itself across a range of staggered ionic energies. While the two-sublattice framework opens up the possibility of investigating magnetic ordering via incorporating some of the modifications to the AA mentioned later in section \ref{aa}, here we choose to focus on the paramagnetic solution at half-filling.

The IHM has received attention as a potential model for describing the ferroelectric transition in perovskite materials~\cite{Egami1} and the neutral-ionic transition in organic mixed-stack donor-acceptor crystals~\cite{Nagaosa1986}. It has been extensively studied in both one dimension (1D) and two dimensions (2D) but the true nature of its phase diagram at half-filling is still undecided. In 1D, Fabrizio~\etal\cite{Fabrizio1} performed a study using bosonization and found that the band gap due to the ionic potential is suppressed at a critical value of $U$ where the system undergoes a transition to an intermediate bond-ordered phase before undergoing a transition to a Mott-insulating phase at a second critical value of $U$. Some contradictory results were obtained, for example Wilkens and Martin~\cite{Wilkens1} found no Mott transition using quantum Monte Carlo (QMC). However, other later studies found results similar to Fabrizio~\etal\cite{Fabrizio1}, for example Batista and Aligia~\cite{Batista1}, Zhang~\etal\cite{Zhang2003} and Manmana~\etal\cite{Manmana1}, the latter two by using density matrix renormalisation group (DMRG) theory.

The situation is far less conclusive in 2D. Using dynamical mean-field theory (DMFT), Garg~\etal\cite{Garg1} found a metallic phase sandwiched between the band insulating phase due to the ionic potential and the Mott insulating phase in the strong coupling regime. Similar results were later obtained by Hoang~\cite{Hoang1} simply by using the CPA. Craco~\etal\cite{Craco1} also used DMFT but found phase transitions of a different nature from metal to Mott insulator and from band insulator to Mott insulator in the phase diagram. Using determinant-QMC, Paris~\etal\cite{Paris1} and Bouadim~\etal\cite{Bouadim1} found that in contrast to the DMFT studies, in the absence of the ionic potential the system is an insulator for any value of $U$ due to the inclusion of intersite magnetic fluctuations. The Mott insulating phase was also found to be anti-ferromagnetic. Subsequently, Byczuk~\etal\cite{Byczuk1} re-examined the IHM using DMFT but allowed for anti-ferromagnetic order and found that the system is insulating for all interaction strengths with no metallic phase. Kancharla and Dagotto~\cite{Kancharla1} used cellular-DMFT to incorporate nonlocal spatial correlations and claimed existence of a bond-ordered intermediate phase instead of a metallic phase, as is the generally accepted situation in 1D. Similar findings were made by Chen~\etal\cite{Chen1} using the variational cluster approach.

This paper is organised as follows. In section~\ref{aa} we introduce the alloy analogy for the IHM and discuss some of the various limitations of the approach. In section \ref{sectioncpa} we explain the CPA solution before showing how to implement for the nonlocal CPA in subsection~\ref{sectionnlcpa}. In section~\ref{results} we present results for the paramagnetic solution of the two-dimensional IHM and compare with previous work, and we conclude in section~\ref{conclusions}.

%--------------------------------------------------------------------------------------------------------------------------------------------
\section{Formalism}\label{formalism}

\subsection{The ``alloy analogy''}\label{aa}

Consider the following Hamiltonian for the single-band IHM on a bipartite lattice with sublattices $A$ and $B$:
\begin{eqnarray}
	H = & \epsilon_{A}\sum_{i\in{A}}n_{i} + \epsilon_{B}\sum_{i\in{B}}n_{i} + U\sum_{i}n_{i\uparrow}n_{i\downarrow} \nonumber\\
        & -t\!\!\! \sum_{i\in{A},j\in{B},\sigma}\!\!\!\left[ c^{+}_{i\sigma} c_{j\sigma}+\textstyle{H.c.}\right]  - \mu\sum_{i}n_{i}
\end{eqnarray}
Here $c^{+}_{i\sigma}\left(c_{i\sigma}\right)$ are the creation (annihilation) operators for an electron with spin $\sigma$ at site $i$, $n_{i\sigma}=c^{+}_{i\sigma}c_{i\sigma}$, and $n_i=n_{i\downarrow}+n_{i\uparrow}$. The nearest-neighbour hopping parameter is denoted by $t$, the on-site Coulomb repulsion by $U$, and the chemical potential by $\mu$. The ionic energies are defined by $\epsilon_{A}=\Delta$ and $\epsilon_{B}=-\Delta$ for sublattices $A$ and $B$ respectively.

For the conventional Hubbard model, Hubbard considered an electron with spin $\sigma$ moving through the lattice and made a static approximation such that the electrons with opposite spin $-\sigma$ were assumed to be fixed at the lattice sites. In other words the dynamics of opposite spin electron populations were treated separately. The electron with spin $\sigma$ could then experience two different types of potential; a potential $U$ at a site which had an electron with spin $-\sigma$ present and a potential $0$ at a site without. Since the values of these potentials were distributed at random, Hubbard viewed the system as a disordered alloy; this is the famous ``alloy analogy" approach to the electron correlation problem mentioned in the introduction. The aim was therefore to solve the alloy problem by approximating the configurational average over all possible disorder configurations. In the case of the IHM the alloy-analogy additionally becomes sublattice-dependent due to the presence of the staggered ionic energies. Following this reasoning we may approximate the above many-body Hamiltonian by the one-electron Hamiltonian
\begin{equation}\label{ham1}
	H = \sum_{i\in{A},\sigma} E_{A\sigma}n_{i\sigma} + \sum_{i\in{B},\sigma}E_{B\sigma}n_{i\sigma} \; -t\!\!\!\sum_{i\in{A},j\in{B},\sigma}\!\!\!\left[ c^{+}_{i\sigma} c_{j\sigma}+\mbox{H.c.}\right]
\end{equation}
where the disorder potential has been defined to include the chemical potential and is given by
\begin{equation}\label{prob}
	E_{\alpha\sigma}=\left\{\begin{array}{l}\epsilon_{\alpha}+{U/2}\;\;\;\mbox{with probability} \;\;\;\langle{n_{\alpha,-\sigma}}\rangle \\  \epsilon_{\alpha}-{U/2}\;\;\;\mbox{with probability}  \;\;\;1-\langle{n_{\alpha,-\sigma}}\rangle \end{array}\right.
\end{equation}
at half-filling with $\mu=U/2$. Here $\langle{n_{\alpha,\sigma}}\rangle$ is the average electron occupancy per site for sublattice $\alpha$ with spin $\sigma$.

It is well known that the AA itself has many shortcomings. These are discussed in the book by Gebhard~\cite{Gebhard1}, and have been found to be mainly due to the arbitrariness in the intuitive way Hubbard performed the decoupling of the higher order Green's functions~\cite{Gebhard1}. Consequently many efforts have been made over the years to improve the approximation. Before going on to develop the formalism for this paper, it is appropriate to mention a small selection of these efforts here.

The first shortcoming to be highlighted is the inability of the AA to describe magnetism~\cite{Brouers1,Brouers2}. For example, in the case of half-filling a transition to an anti-ferromagnetic Mott insulator is not observed in the strong-coupling regime and it is only possible to obtain the paramagnetic solution. One of the better known theories aimed at correcting the AA in the strong-coupling limit is the modified alloy-analogy (MAA) of Herrmann and Nolting~\cite{Herrmann1}, which was derived by combining the spectral density approach~\cite{Geipel1,Nolting1} (SDA) with the CPA in order to reproduce the exact results of Harris and Lange~\cite{Harris1} as $U\rightarrow\infty$. The MAA is based upon the notion that the atomic levels and concentrations used in Hubbard's AA are not the only choice available for constructing an alloy analogy. The MAA still uses the CPA equation but replaces the atomic levels and concentrations with modified expressions which include a band-shift correction. Potthoff~\etal\cite{Potthoff1} showed that this choice is optimal for reproducing the first four moments of the spectral density in contrast to the first three in Hubbard's AA. Unlike Hubbard's AA, the MAA is capable of producing magnetic solutions, albeit over restricted parameter ranges. Another approach capable of capturing the transition to anti-ferromagnetic Mott insulator is the mean-field theory of Janis and Vollhardt~\cite{Janis1,Janis2} who constructed a thermodynamically-consistent alloy analogy.

A second deficiency of the AA is that the metallic state is not described as a Fermi liquid. This is a natural consequence of viewing the system in terms of a set of static impurity scattering centres which leads to a self-energy with a finite imaginary part at the Fermi level. Although including the resonance broadening correction of the full Hubbard III approximation will take into account dynamic processes at the impurity scattering centre by relaxing the assumption that the opposite spin electrons are fixed at the lattice sites, this is still not a true description of dynamics in the sense that the effective field seen by the spin $\sigma$ electron is not time-dependent. Nevertheless, Edwards and Hertz~\cite{Edwards1,Edwards2} used the limit of infinite dimensions to contruct a method for correcting this non-Fermi liquid behaviour in the weak-coupling limit using a modified Green's function, the method being exact for small $U$ up to order $U^2$. Furthermore, since the MAA and the Edwards and Hertz approximation (EHA) correct opposite extremes of the AA, Potthoff~\etal\cite{Potthoff1} subsequently developed the interpolating alloy-analogy (IAA) which suitably combines the MAA and EHA into one computational scheme which retains the advantages and avoids the deficiencies of the respective approaches.

A further means of modifying the AA was investigated by Corrias~\cite{Corrias1,Corrias2} who took into account a third type of site not addressed by Hubbard when considering the motion of a spin $\sigma$ electron, specifically the sites already occupied by other spin $\sigma$ electrons to which hopping is forbidden by the Pauli principle. Modification of the Green's function decoupling performed by Hubbard is another avenue for improvement, and such methods include the correlation Green's function method of Luo and Wang~\cite{Luo1} which adds bandwidth and band-shift higher-order correlation effects, and the recent work of Gorski and Mizia~\cite{Gorski1,Gorski2}.

%--------------------------------------------------------------------------------------------------------------------------------------------
\subsection{CPA for the ionic Hubbard model}\label{sectioncpa}

The IHM has previously been studied using the CPA by Hoang~\cite{Hoang1}. This in turn is based on the earlier framework of Gupta~\etal\cite{Gupta1} who investigated the possibility of magnetic ordering in the Hubbard model using a two-sublattice version of the CPA which we refer to as the ``coupled-CPA'' method.

In principle, the Green's function corresponding to the Hamiltonian of equation (\ref{ham1}) needs to be averaged over all possible disorder configurations. Such an exact average Green's function would define an effective medium describing exactly the average properties of a single electron. Since such an exact average is not feasible, the CPA introduces a simplified effective medium corresponding to the approximate Hamiltonian
\begin{equation}\label{hcpa}
	\fl H_{CPA} = \sum_{i\in{A},\sigma} \Sigma_{A\sigma}n_{i\sigma} + \sum_{i\in{B},\sigma}\Sigma_{B\sigma}n_{i\sigma}\; -t\!\!\!\sum_{i\in{A},j\in{B},\sigma}\!\!\!\left[ c^{+}_{i\sigma} c_{j\sigma}+\mbox{H.c.}\right]
\end{equation}
where each sublattice $A(B)$ has its own single-site or local self-energy $\Sigma_{A\sigma(B\sigma)}$ situated at every sublattice site. In $\b{k}$-space, the CPA average Green's function can be written in the two-sublattice matrix form
\begin{equation}
	\bar{\u{G}}_{\sigma}(\b{k},\omega)=\left[ \begin{array}{cc} \omega-\Sigma_{A\sigma}(\omega) & -t(\b{k}) \\ -t(\b{k}) & \omega-\Sigma_{B\sigma}(\omega) \end{array}\right]^{-1}.
\end{equation}
Carrying out the matrix inversion yields
\begin{equation}
	\bar{G}_{AA\sigma}(\b{k},\omega)=\frac{\omega-\Sigma_{B\sigma}}{(\omega-\Sigma_{A\sigma})(\omega-\Sigma_{B\sigma})-t(\b{k})^{2}}
\end{equation}
\begin{equation}
	\bar{G}_{BB\sigma}(\b{k},\omega)=\frac{\omega-\Sigma_{A\sigma}}{(\omega-\Sigma_{A\sigma})(\omega-\Sigma_{B\sigma})-t(\b{k})^{2}}
\end{equation}
for the sites on sublattice $A$ and $B$ respectively, where $\b{k}$ belongs to the first BZ of the sublattice considered. In real space we have
\begin{equation}\label{cpasc1}
	\bar{G}_{\alpha\sigma}(\omega)=\frac{1}{\Omega_{BZ}}\int_{\Omega_{BZ}}d{\b{k}}\;\frac{\omega-\Sigma_{\bar{\alpha}\sigma}}{(\omega-\Sigma_{A\sigma})(\omega-\Sigma_{B\sigma})-t(\b{k})^{2}}
\end{equation}
where $\alpha=A(B)$, $\bar{\alpha}=B(A)$ and the integral is over the first BZ of the sublattice. Since we are only considering the diagonal terms of the Green's function matrix, notation has been adopted such that $\alpha\equiv\alpha\alpha$ for clarity. Using the relation for the bare integrated DOS
\begin{equation}
	\int{dE}\;\rho_{0}(E)=\int{dE}\;\frac{1}{N}\sum_{\b{k}}\delta(E-t(\b{k}))
\end{equation}
for real energies $E$, equation (\ref{cpasc1}) can be re-written in the form
\begin{equation}
	G_{\alpha}(\omega)=(\omega-\Sigma_{\bar{\alpha}})\int{dE}\;\frac{\rho_{0}(E)}{(\omega-\Sigma_{A})(\omega-\Sigma_{B})-E^{2}}
\end{equation}
for which approximations to the bare DOS can be made. To determine the CPA self-energies, Hoang~\cite{Hoang1} proceeded by employing Hubbard's semi-elliptic model DOS~\cite{Hubbard1}. This simplifies the above integral and yields a pair of equations for $\Sigma_{A\sigma}$ and $\Sigma_{B\sigma}$ which need to be solved self-consistently. Here we do not make any approximations for the DOS; instead we determine the self-energies numerically. We begin by defining the cavity Green's function $\cal{G}_{\alpha\sigma}(\omega)$ through the relation
\begin{equation}
	{\cal{G}}_{\alpha\sigma}^{-1}(\omega)=\bar{G}_{\alpha\sigma}^{-1}(\omega)+\Sigma_{\alpha\sigma}(\omega)
\end{equation}
for each sublattice $\alpha$, which describes the medium with the self-energy at some chosen site on each sublattice removed i.e.~a cavity. In this definition it is not necessary to consider the off-diagonal terms in the Green's function matrix, $\bar{G}_{AB\sigma}$ and $\bar{G}_{BA\sigma}$, since there are no self-energy terms coupling the sublattices. We may now fill the cavity on each sublattice with some real ``impurity'' configuration by defining the impurity Green's functions
\begin{equation}
  G^{\gamma}_{\alpha\sigma}(\omega)=\left[{\cal{G}}_{\alpha\sigma}^{-1}(\omega)-E^{\gamma}_{\alpha\sigma}\right]^{-1}
\end{equation}
with impurity configurations $E^{\gamma=\pm}_{\alpha\sigma}=\epsilon_{\alpha}{\pm}U/2$ as defined by equation (\ref{prob}). The CPA demands that
\begin{equation}\label{cpasc2}
	\langle{G^{\gamma}_{\alpha\sigma}(\omega)}\rangle=\bar{G}_{\alpha\sigma}(\omega),
\end{equation}
where the average is taken over the impurity configuration probablities defined by equation (\ref{prob}), i.e.
\begin{equation}
	\langle{n_{\alpha,-\sigma}}\rangle G^{\gamma=+}_{\alpha\sigma}(\omega) + \left(1-\langle{n_{\alpha,-\sigma}}\rangle\right) G^{\gamma=-}_{\alpha\sigma}(\omega)  =\bar{G}_{\alpha\sigma}(\omega).
\end{equation}
Equations (\ref{cpasc1}) and (\ref{cpasc2}) thus need to be solved self-consistently. Since in the IHM electrons generally prefer to be on sublattice $B$ and we have $\langle{n_{A}}\rangle+\langle{n_{B}}\rangle=2$ at half-filling where
\begin{equation}
	\langle{n_{\alpha}}\rangle=\langle{n_{\alpha\uparrow}}\rangle+\langle{n_{\alpha\downarrow}}\rangle=-\frac{1}{\pi}\int_{-\infty}^{0}\mbox{Im}\left[{\bar{G}_{\alpha\uparrow}+\bar{G}_{\alpha\downarrow}}\right]d\omega,
\end{equation}
it is also necessary to ensure the resulting integrated DOS for each sublattice are both consistent with the average occupation number probabilities used in equation~(\ref{cpasc2}), thus adding an extra layer of self-consistency.

As mentioned in the introduction, there are no anti-ferromagnetic or ferromagnetic CPA solutions for the (non-degenerate) Hubbard model~\cite{Brouers1,Brouers2}. This is also true in the presence of staggered ionic energies and so here we are restricted to the paramagnetic solution. Thus the Green's function is the same for both spin populations so that $G_{\alpha\uparrow} = G_{\alpha\downarrow}$ and $\langle{n_{\alpha\uparrow}}\rangle = \langle{n_{\alpha\downarrow}}\rangle =  \langle{n_{\alpha}}\rangle/2$. Finally note that although there are no off-diagonal self-energy terms coupling the sublattices, equation (\ref{cpasc1}) shows that the sublattice self-energies are not determined independently of each other.

%--------------------------------------------------------------------------------------------------------------------------------------------
\subsection{Nonlocal CPA for the ionic Hubbard model}\label{sectionnlcpa}

The nonlocal CPA introduced by Jarrell and Krishnamurthy~\cite{Jarrell1} systematically includes nonlocal correlations beyond the CPA via the self-consistent embedding of a cluster of sites with periodic boundary conditions imposed. The reader is referred to references~\cite{Jarrell1,Rowlands1} for full details of the method; here we simply show how to implement for the IHM.

\begin{figure}
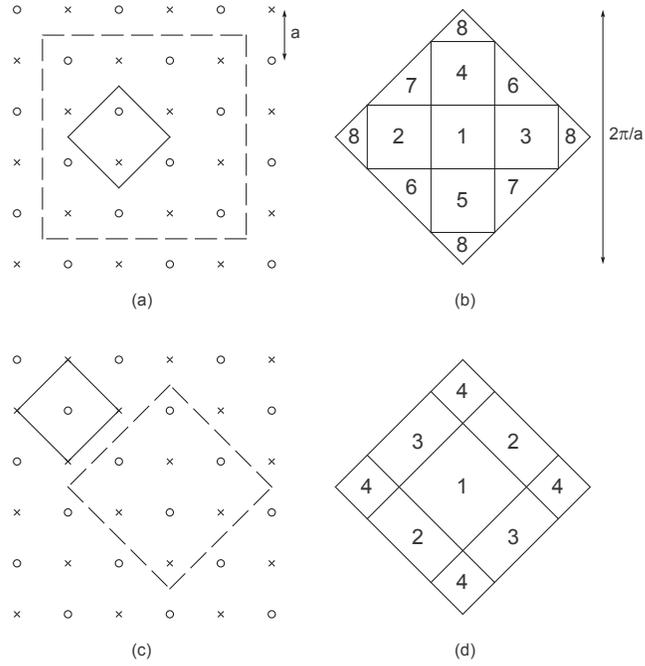

 \begin{center}
 \scalebox{0.8}{\i{cluster}}
 \caption{\label{cluster}(a) $N_c=16$ real-space cluster for the bipartite lattice indicated by the dashed square. Sublattice $A$ sites are denoted by circles and sublattice $B$ sites are denoted by crosses. The lattice unit cell is shown by the solid polygon. The nearest neighbour lattice constant is $a$. (b) Brillouin zone for one sublattice divided into 8 regions each centred at a cluster momentum point $\b{K}_n$. The height and width of the sublattice Brillouin zone is $2\pi/a$. Two sublattice Brillouin zones together form 16 regions of $\b{k}$-space for the $N_c=16$ real-space cluster. (c) and (d): Same as (a) and (b) but for the $N_c=8$ cluster.}
 \end{center}
\end{figure}

Because of the boundary conditions, the mapping of the cluster to the lattice is achieved in $\b{k}$-space~\cite{Jarrell1}. The BZ is divided into $N_c$ regions centred at the set of  ``cluster momenta'' $\{\b{K}\}$. Correspondingly in real space the lattice is divided into $N_c$ sublattices, and a site from each sublattice taken together form a real-space cluster~\cite{Rowlands1}. Some example real space clusters and corresponding BZ regions for the 2D bipartite lattice are illustrated in figure \ref{cluster}. The main approximation made is to represent the exact lattice self-energy in $\b{k}$-space $\Sigma(\b{k},\omega)$ by the step function $\Sigma(\b{K},\omega)$ in the regions centred at the cluster momenta. In this approximation the translational invariance of the underlying lattice is always preserved irrespective of the size of the cluster. In the case of the IHM, the translational invariance of the underlying two-sublattice unit cell must be preserved and hence $\Sigma(\b{K},\omega)$ also becomes sublattice-dependent. Therefore an effective medium is introduced corresponding to the approximate $\b{k}$-space Hamiltonian
\begin{eqnarray}\label{hnlcpa}
	H &=& \sum_{\b{K},\b{k'}}\Sigma_{AA}(\b{K},\omega)n_{A,\b{K}+\b{k'}} + \sum_{\b{K},\b{k'}}\Sigma_{BB}(\b{K},\omega)n_{B,\b{K}+\b{k'}} \nonumber\\
     & & -\sum_{\b{K},\b{k'}}\left[\left(t(\b{K}+\b{k'})+\Sigma_{AB}(\b{K},\omega)\right) c^{+}_{A,\b{K}+\b{k'}} \,\, c^{}_{B,\b{K}+\b{k'}}+\mbox{H.c.}\right]
\end{eqnarray}
where all spin indices have been omitted for clarity. The corresponding Green's function matrix is defined by
\begin{equation}\label{det}
	\fl\bar{\u{G}}(\b{K},\b{k'},\omega) = \left[ \begin{array}{cc} \omega-\Sigma_{AA}(\b{K},\omega) & -t(\b{K}\!+\!\b{k'})-\Sigma_{AB}(\b{K},\omega) \\ -t(\b{K}\!+\!\b{k'})-\Sigma_{BA}(\b{K},\omega) & \omega-\Sigma_{BB}(\b{K},\omega) \end{array} \right]^{-1}
\end{equation}
where all $\b{K}$ and $\b{k'}$ belong to the first BZ of the two-sublattice unit cell. This matrix is similar to that of the CPA except that there are terms $\Sigma_{AB}$ and $\Sigma_{BA}$ coupling the sublattices and all self-energies are momentum-dependent. Carrying out the matrix inversion yields
\begin{equation}
	\fl\bar{\u{G}}(\b{K},\b{k'},\omega) = \frac{1}{f(\b{K},\b{k'})} \left[ \begin{array}{cc} \omega-\Sigma_{BB}(\b{K},\omega) & t(\b{K}\!+\!\b{k'})+\Sigma_{AB}(\b{K},\omega) \\
  t(\b{K}\!+\!\b{k'})+\Sigma_{BA}(\b{K},\omega) & \omega-\Sigma_{AA}(\b{K},\omega) \end{array} \right]
\end{equation}
where $f$ is the determinant of the matrix in equation (\ref{det}) and is given by
\begin{eqnarray}
	f(\b{K},\b{k'}) = &  (\omega-\Sigma_{AA}(\b{K}))(\omega-\Sigma_{BB}(\b{K})) \nonumber\\
                              & -(t(\b{K}\!+\!\b{k'})\!+\!\Sigma_{AB}(\b{K})) (t(\b{K}\!+\!\b{k'})\!+\!\Sigma_{BA}(\b{K}))
\end{eqnarray}
The next step is to define the ``coarse-grained'' $\b{k}$-space sublattice Green's functions by integrating over each region of $\b{k}$-space or ``tile'' $n$ centred at the cluster momenta $\b{K}$:
\begin{equation}
  \bar{G}_{\alpha\beta}(\b{K}_n,\omega)=\frac{N_c}{2}\frac{1}{\Omega_{BZ}}\int_{\Omega_{\b{K}_n}} d\b{k'}\; \bar{G}_{\alpha\beta}(\b{K},\b{k'},\omega)
\end{equation}
These can be transformed to real space using the Fourier transform
\begin{equation}\label{nlcpasc1}
  \bar{G}_{\alpha\beta}^{IJ}(\omega)=\frac{2}{N_c}\sum_{\b{K}_n}\; \bar{G}_{\alpha\beta}(\b{K}_n,\omega) e^{i\b{K}_n(\b{R}_I-\b{R}_J)}
\end{equation}
for cluster matrix elements $\{I,J\}$ where $N_c$ is the number of sites in the cluster. Together these form the supermatrix
\begin{equation}
	\bar{\u{\u{G}}}(\omega) = \left[ \begin{array}{cc} \bar{\u{G}}_{AA}(\omega) &  \bar{\u{G}}_{AB}(\omega) \\  \bar{\u{G}}_{BA}(\omega) &  \bar{\u{G}}_{BB}(\omega) \end{array} \right]
\end{equation}
which has dimension $N_c$. It can be seen that sites belonging to sublattices $A$ and $B$ are grouped separately. We can similarly Fourier-transform the $\b{k}$-space self-energies to real space to obtain the cluster self-energy matrix $\u{\u{\Sigma}}$. To determine the Green's functions and self-energies, we define the cavity Green's function $\cal{\u{\u{G}}}(\omega)$ through the relation
\begin{equation}\label{cavnlcpa}
	{\cal{\u{\u{G}}}}^{-1}(\omega)=\bar{\u{\u{G}}}^{-1}(\omega)+\u{\u{\Sigma}}(\omega).
\end{equation}
Note that unlike the ``coupled-CPA'' method of section \ref{sectioncpa}, here it is necessary to include the inter-sublattice terms in the Green's function matrix, $\bar{\u{G}}_{AB}$ and $\bar{\u{G}}_{BA}$, when defining the cavity Green's function due to the presence of the inter-sublattice self-energy terms. The impurity Green's function is then given by
\begin{equation}
  \u{\u{G}}^{\gamma}(\omega)=\left[{\cal{\u{\u{G}}}}^{-1}(\omega)-\u{\u{E}}^{\gamma}\right]^{-1}
\end{equation}
for some cluster disorder configuration $\gamma$, where the site component of each configuration is one of $\epsilon_{A}{\pm}U/2$ or one of  $\epsilon_{B}{\pm}U/2$ as defined by equation (\ref{prob}) for every site on sublattice $A$ and $B$. The nonlocal CPA demands that
\begin{equation}\label{nlcpasc2}
	\langle{\u{\u{G}}^{\gamma}(\omega)}\rangle=\bar{\u{\u{G}}}(\omega)
\end{equation}
where the average is taken over all $2^{N_c}$ possible cluster configurations with the site probability component of each cluster probability defined by equation (\ref{prob}). The nonlocal CPA solution is obtained by solving equations (\ref{nlcpasc1}) and (\ref{nlcpasc2}) self-consistently. Again it is necessary to ensure that the resulting integrated DOS for each cluster site is consistent with the relevant site occupation number probability used in the cluster configurations of equation~(\ref{nlcpasc2}),
thus adding an extra layer of self-consistency. As a consequence of translational invariance, all sites on sublattice $A$ will have the same average occupation number probability (integrated DOS per site), as will all sites on sublattice $B$. At half-filling, the integrated DOS summed over all cluster sites should equal $N_c$.

In the limit where $\Delta=0$, the two sublattices are equivalent. Nevertheless, the system can still be viewed in terms of a two-sublattice unit cell. Then for a cluster of size $N_c$ it is still only necessary to consider the $N_c/2$ cluster momenta associated with the BZ of the two-sublattice unit cell. This is because for $\Delta=0$ the BZ integration with $N_c$ cluster momenta for the usual single-site unit cell becomes equivalent to two sublattice-BZ integrations, each with $N_c/2$ cluster momenta, which in turn are equivalent to the BZ of the two-sublattice unit cell. It is also worth looking at the limit where the cluster size $N_c=2$. In that case there will still be a different self-energy value for each sublattice in $\b{k}$-space and hence off-diagonal self-energy terms which couple the two sublattices in real space. Hence we obtain an improved version of the coupled-CPA method of section \ref{sectioncpa}. It should also be mentioned that the choice made for the set of cluster momenta is not unique, leading to non-unique results for small cluster sizes~\cite{Rowlands2}. A reformulation of the nonlocal CPA which uses multiple sets of cluster momenta has been derived which corrects this problem~\cite{Rowlands2}, also naturally providing a lattice self-energy and Green's function unlike equations (\ref{nlcpasc1}) and (\ref{nlcpasc2}) which strictly speaking are cluster quantities. However, in this manuscript we use only one set of cluster momenta for simplicity.

Finally, we note that a multi-sublattice implementation of the nonlocal CPA has previously been carried out by Marmodoro and Staunton for a 1D tight-binding model alloy~\cite{Marmodoro1}. Our formulation for the IHM above can be viewed as an important extension to the case of strongly-correlated systems and has potential applications to various strongly-correlated models like the d-p model originally proposed for describing high-T$_c$ cuprates~\cite{Emery1987}, the Hubbard model on the honeycomb lattice~\cite{Meng2010} and the extended Hubbard model on the checkerboard lattice~\cite{Zhang2005}, etc. 

%--------------------------------------------------------------------------------------------------------------------------------------------
\section{Results}\label{results}

\begin{figure}
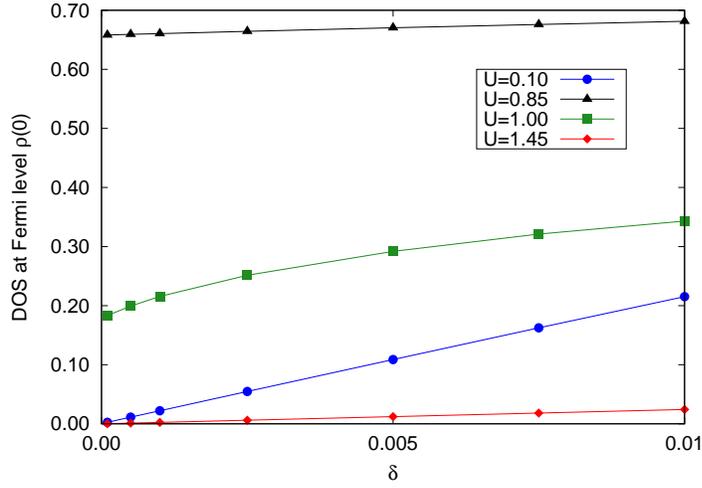

 \begin{center}
 \scalebox{0.75}{\i{def}}
 \caption{\label{def}DOS at the Fermi level $\rho(0)$ as a function of imaginary energy part $\delta$ for ionic energy $\Delta=0.10$ and a selection of on-site interaction values $U$. Extrapolation to $\delta=0$ via polynomial fitting indicates that $U$=0.85 and $U$=1.00 represent metallic phases while $U=0.10$ and $U=1.45$ represent insulating phases.}
 \end{center}
\end{figure}

\begin{figure}
 \begin{center}
 \scalebox{0.75}{\i{phase_nc4}}
 \caption{\label{phase4}$T=0$ phase diagram for the IHM obtained using the CPA (dashed blue line) and nonlocal CPA with $N_c=4$ (solid line). MI, M, and BI denote Mott insulator, metal, and band insulator, respectively. The small shaded patterned region represents a spurious insulating phase within the metallic sector as described in the text.}
 \end{center}
\end{figure}

\begin{figure}
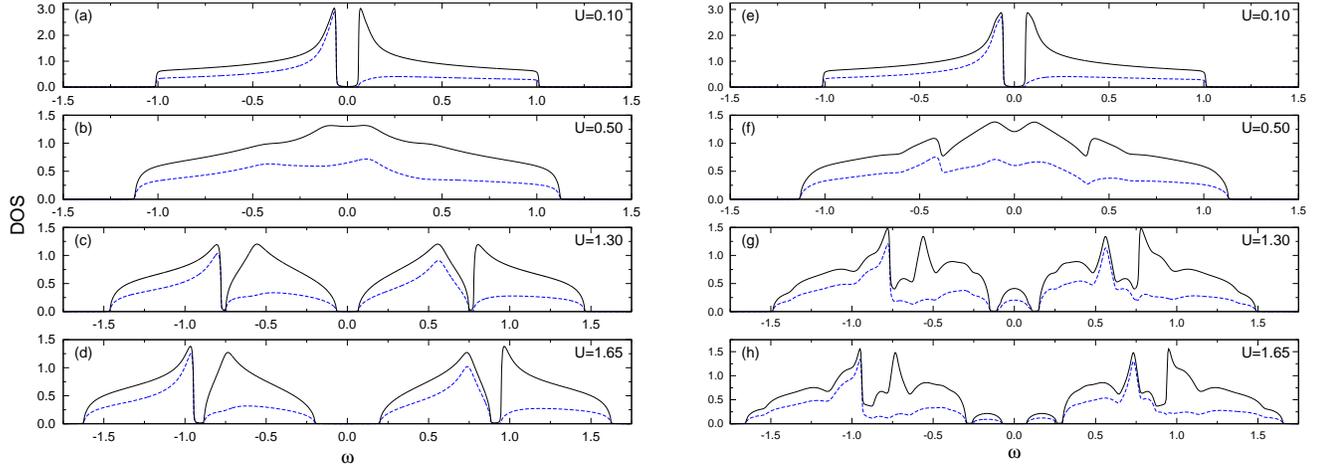

 \begin{center}
 \begin{tabular}{ll}
 \scalebox{0.70}{\i{dos_cpa}} &  \scalebox{0.70}{\i{dos_nlcpa_nc4}}  \\
 \end{tabular}
 \caption{\label{dos4}(a)-(d) Total DOS per two-site unit cell (solid line) and local DOS for the $B$ sublattice (dashed line) for $\Delta=0.1$ and various values of $U$ calculated using the CPA. The results shown are for one spin direction only so that integrating the total DOS up to the Fermi level yields one electron. The energy imaginary part (smoothing factor) is $\delta$=0.001. (e)-(h) Same but calculated using the nonlocal CPA with $N_c=4$.}
 \end{center}
\end{figure}

\begin{figure}
 \begin{center}
 \scalebox{0.75}{\i{phase_nc8}}
 \caption{\label{phase8}$T=0$ phase diagram for the IHM obtained using the CPA (dashed blue line) and nonlocal CPA with $N_c=8$ (solid lines). MI, M, and BI denote Mott insulator, metal, and band insulator, respectively. The two shaded patterned regions are spurious insulating phases within the metallic sector as described in the text.}
 \end{center}
\end{figure}

\begin{figure}
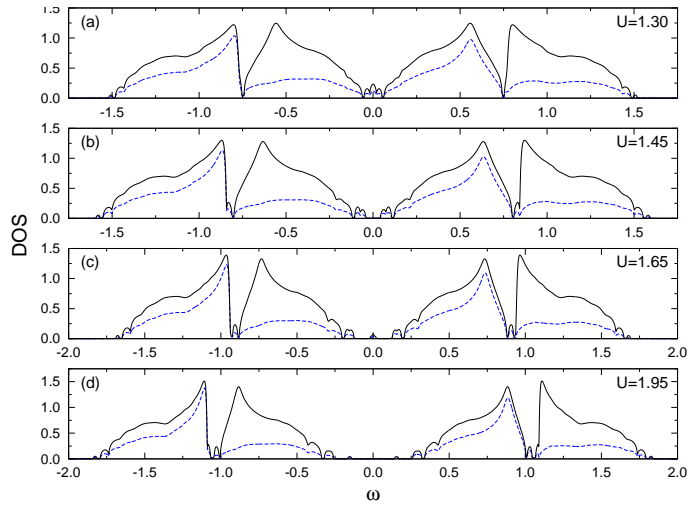

 \begin{center}
 \scalebox{0.75}{\i{dos_nlcpa_nc8}}
 \caption{\label{dos8}Total DOS per two-site unit cell (solid line) and local DOS for the $B$ sublattice (dashed line) for $\Delta=0.1$ and various values of $U$ calculated using the nonlocal CPA with $N_c=8$. Notice the very small peak at the Fermi level in figure (c) which has split into two in figure (d).}
 \end{center}
\end{figure}

\begin{figure}
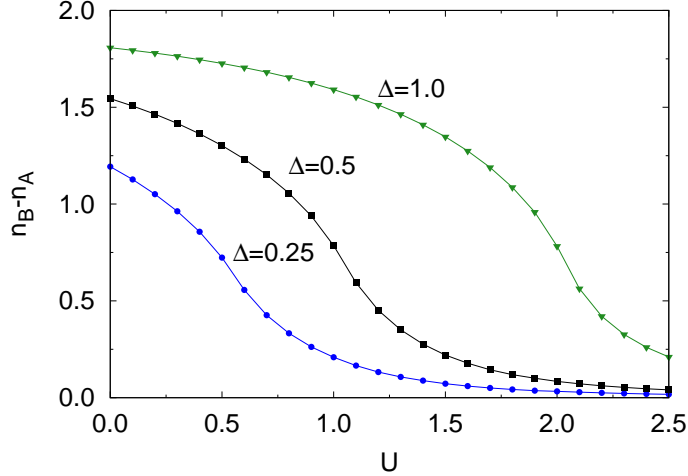

 \begin{center}
 \scalebox{0.75}{\i{charge_nc8}}
 \caption{\label{charge}(a) Staggered charge density $n_B-n_A$ as a function of $U$ for different values of $\Delta$ using the nonlocal CPA with $N_c=8$. Notice the change in slope of each curve just after $U=2\Delta$ which indicates the band insulator to metal transition.}
 \end{center}
\end{figure}

The 2D phase diagram for the system at zero temperature has previously been calculated using the CPA by Hoang~\cite{Hoang1} who compared the CPA results with those given by other approaches. Since our main motivation here is to study the effect of intersite correlations using the nonlocal CPA, we calculate the 2D phase diagram for the system using the same parameters as Hoang~\cite{Hoang1} so that direct comparisons can be made. Thus the energy unit is set by $W=4t=1$ where $W$ is the half-width of the band. In order to determine the phase boundaries, the CPA or nonlocal CPA medium and site occupation number probabilities were first determined self-consistently using energies with an imaginary part $\delta=0.01$. Using the same number probabilities, the DOS at the Fermi level $\rho(0)$ was then calculated by determining the medium self-consistently as a function of $\delta$. Extrapolation via polynomial fitting was used in order to obtain the value with $\delta\rightarrow{0}^{+}$. In the case of a finite gap, $\rho(0)$ extrapolates to zero. In contrast $\rho(0)$ remains finite in the metallic region~\cite{Craco1}. Taking into account tolerances used in the calculations, $\rho(0)\ge{0.01}$ was considered to be a finite value. Figure \ref{def} shows example CPA calculations for $\rho(0)$ as a function of $\delta$. It is quite clear that the graphs for $U=0.85$ and $U=1.00$ will not extrapolate to zero via polynomial fitting and hence indicate metallic phases, whilst the graphs for $U=0.1$ and $U=1.45$ indicate insulating phases.

The phase diagram for the system calculated using both the CPA and nonlocal CPA with a cluster size $N_c=4$ is shown in figure~\ref{phase4}, with the CPA phase boundaries denoted by dotted lines and nonlocal CPA by solid lines. The CPA phase diagram is similar to that of Hoang~\cite{Hoang1}, though the metallic phase is very slightly wider. The small difference in shape in the weak-coupling regime is likely to be due to the fact that a model DOS has not been used in our calculations. It can be seen that the phase boundary between the band insulator and metallic phases is unchanged in the nonlocal CPA calculations compared to the CPA and approaches the line $U=2\Delta$. On the other hand, the phase boundary for the MIT is significantly raised from the CPA value in the weak-coupling regime, but again approaches the strong-coupling line $U=2\Delta$ as $\Delta$ is increased. Thus intersite spatial correlations appear to enhance the metallic phase compared to the CPA, particularly in the weak-coupling regime. We note that for $\Delta=0$ this behaviour is in agreement with the results of both Mors~\etal\cite{Mors1} and Luo~\etal\cite{Luo2} for the conventional Hubbard model, where according to their methods it was found that intersite spatial correlations shift the critical value of $U$ for the MIT to a higher value. Also present in figure~\ref{phase4} is a small insulating region within the metallic sector between $\Delta=0.0$ and $\Delta=0.25$ which we believe to be unphysical and resulting from a shortcoming of the nonlocal CPA method for small cluster sizes (see $N_c=8$ calculation below).

To identify the origin of the differences between the CPA and nonlocal CPA phase diagrams, we now examine some example DOS results. First we consider the CPA. Figure \ref{dos4} (a)-(d) show plots with ionic energy $\Delta=0.1$ for a selection of values of $U$. The general features of the plots are similar to those of Hoang~\cite{Hoang1}, though the curves are of more realistic shape since here the use of a model DOS has been avoided. For $U=0.1$ the system is a band insulator with charge gap slightly smaller than the value $2\Delta$ for the non-interacting ($U=0)$ limit. The results for values $U=0.50$ and $U=1.30$ clearly lie within the metallic and Mott insulating regimes respectively. The critical value for the metal-insulator transition is found to be very close to $U=1$.

The nonlocal CPA results for the same parameters with $N_c=4$ are shown in figure \ref{dos4} (e)-(h). The curve for $U=0.1$ is similar to that of the CPA, indicating that for relatively small values of $U$ the effect of intersite correlations is small. On the other hand, immediately apparent is the emergence of a peak at the Fermi level in the metallic phase as seen in the example plot for $U=1.30$. As the Coulomb repulsion is increased further, the peak gradually diminishes. However, the transition from metal to Mott insulator is significantly delayed resulting in a larger critical value for $U$. A similar peak for the conventional Hubbard model (i.e.~$\Delta=0$) was found by Luo~\etal\cite{Luo2} using their method involving combination of the SDA with the full Hubbard III approximation and making a nonlocal self-energy ansatz. In their work it was interpreted as a quasiparticle peak resulting from the nonlocal nature of the self-energy i.e.~a different mechanism to the well-known peak observed in DMFT calculations due to the dynamical though local nature of the self-energy. However Luo~\etal\cite{Luo2} do not indicate the nature of the imaginary part of the self-energy at the Fermi level. In fact a finite value was observed in DMFT calculations~\cite{Craco1} which implies a finite lifetime for the state. Craco~\etal\cite{Craco1} identified this as being a consequence of a pseudo-gap feature at the Fermi level in the DOS on the $B$-sublattice, meaning that the one-electron DOS in the metallic phase is not attached to the unperturbed ($U=\Delta=0$) value. In our work we also obtain a finite value but we regard this as being in agreement with the notion that viewing the system in terms of a set of static impurity scattering centres in the AA approach rules out the possibility of a Fermi liquid. Hence we consider the presence of the peak to be simply due to DOS at the band edges arising from cluster contributions as is the case in conventional disordered alloy calculations in the split-band regime.

In order to study the effect of longer-ranged intersite correlations, the nonlocal CPA phase diagram calculated using a larger cluster size, $N_c=8$, is shown in figure \ref{phase8} together with the earlier CPA calculation. The lower phase boundary between band insulator and metal shows similar behaviour to the $N_c=4$ calculation. As expected, the small spurious insulating region present in the $N_c=4$ phase diagram for values of $U$ just below the upper CPA phase boundary has disappeared. On the other hand, as $U$ is increased above the upper CPA phase boundary the system surprisingly appears to switch back and forth between metal and Mott insulator so that the MIT occurs three times. We believe the reason for this lies in a shortcoming of the conventional nonlocal CPA method. The medium self-energy should be a smooth function in $\b{k}$-space for all cluster sizes. However, for the purposes of producing an analytic Green's function, the nonlocal CPA approximates the medium self-energy as a step function in $\b{k}$-space~\cite{Jarrell1}. It has previously been shown that such an approximation can lead to spurious DOS results for small cluster sizes where the jumps between the steps are large~\cite{Rowlands6}. In particular, artificial gaps appear at certain energies. In the present context, this could cause artificial gaps to appear at the Fermi level for certain values of $U$, resulting the extra insulating phases denoted by the shaded regions in the phase diagrams of figures \ref{phase4} and \ref{phase8}. This can be seen in the example DOS plots of figure \ref{dos8} for cluster size $N_c=8$, where after undergoing a MIT between $U=1.30$ and $U=1.45$, a small peak appears again at the Fermi level around $U=1.65$ which subsequently separates into two very small peaks as shown for $U=1.95$ where they are well separated. These small peaks are contributions to the DOS at the very edges of the bands from cluster configurations. We therefore believe that the gap seen at the Fermi level for $U=1.45$ is an artificial gap and the final MIT just after $U=1.65$ is the true one because only the smaller critical $U$ values can result from overlap of bands with the spurious gaps. This is also in agreement with a systematic enhancement of the metallic phase with increasing cluster size.

We point out that a reformulation of the nonlocal CPA has been derived which systematically corrects the self-energy step function approximation by producing a smoother function in $\b{k}$-space for any given cluster size~\cite{Rowlands2}. This is achieved by utilising the fact that the choice of cluster momenta for a given cluster size is not unique. By using multiple sets of cluster momenta, a smoother and more realistic $\b{k}$-space self-energy can be self-consistently determined by reducing the jumps between the steps. It was found that implementing this method removes the spurious gaps in the DOS for disordered alloy calculations~\cite{Rowlands2}. It should be stressed that this reformulation simply produces the correct self-energy curve and DOS for a given cluster size and hence correlation length; information beyond the correlation length can only be introduced by using a larger cluster size. In the present context of the IHM, we believe that implementing the method would remove the spurious extra phase boundaries present in figures \ref{phase4} and \ref{phase8}. This is left as future work.

In order to help identify the nature of the phase transitions in figure \ref{phase8}, the staggered average occupation number (charge density) $n_B-n_A$ as a function of $U$ for different values of $\Delta$ using the nonlocal CPA with $N_c=8$ is shown in figure \ref{charge}.  A charge density wave is present throughout the whole parameter regime and gradually softens with increasing $U$. We find that the difference between the nonlocal CPA and the CPA results are negligible, of order 5$e$-03, and the phase transitions are seen to be continuous in all cases. Our results are in excellent agreement with the CPA results of Hoang~\cite{Hoang1}, which in turn are in good agreement with those obtained by single-site DMFT~\cite{Byczuk1}. Particularly noticeable for all three $\Delta$ values in our results are the change in slopes just after $U=2\Delta$. These indicate the band insulator to metal transition which is dominated by the charge density wave. On the other hand, the MIT is governed by the Hubbard $U$ and so the curves smoothly change with no inflection point at the MIT since they have no relation to the charge density wave.

Since cellular-DMFT calculations for the IHM in 2D indicated an intermediate bond-ordered phase~\cite{Kancharla1}, it should be checked that the intermediate phase found in our calculations is indeed metallic. A particularly advantageous feature of the nonlocal CPA in the context of conventional alloy calculations is that the medium preserves the translational invariance of the underlying lattice. Unfortunately, in the present context this rules out the possibility of a long-range bond-ordered phase by construction. Therefore in order to determine whether intersite spatial correlations are able to produce a long-range bond ordered phase within the AA approach, we also implemented the molecular CPA~\cite{Tsukada1,Ducastelle2} for the IHM which naturally describes long-range order up to the length scale of a ``molecule'' or cluster of sites. Indeed, cellular-DMFT is the analog of the molecular CPA for correlated electron systems. The implementation of the molecular CPA for the IHM is described in the Appendix. The bonds are given by the matrix elements $\{\bar{G}^{IJ}\}$ for nearest neighbour sites $I,J$ within the cluster. We performed extensive calculations in 1D and found no evidence of a long-range bond-ordered phase within the AA approach.

Finally, the present calculations could be improved by using some of the modifications described in section \ref{aa} such as the interpolating alloy-analogy (IAA)~\cite{Potthoff1}. For the paramagnetic solution at half-filling, the IAA reduces to the conventional AA in the strong-coupling regime so it would only be necessary to correct the non-Fermi liquid behaviour in the weak-coupling regime using the Edwards-Hertz approximation. A further improvement would be to include Hubbard's resonance broadening correction and hence generalise the full Hubbard III approximation.

%--------------------------------------------------------------------------------------------------------------------------------------------
\section{Conclusions}\label{conclusions}

We have obtained the paramagnetic solution of the nonlocal CPA for the two-dimensional IHM at half-filling and zero temperature. We found that intersite (nonlocal) spatial correlations enhance the intermediate metallic phase throughout the range of parameters, although the difference decreases with increasing $\Delta$. While the critical value $U_{c1}$ for the transition from band insulator to metal is very similar to that of the CPA, a peak is developed in the DOS in the metallic phase as $U$ is increased which delays the transition from metal to Mott insulator, siginificantly raising the critical value $U_{c2}$ for the MIT. To confirm the metallic nature of the intermediate phase, we also implemented the molecular CPA for the IHM and found no evidence of an intermediate bond-ordered phase. These findings further support the view~\cite{Garg1,Hoang1,Craco1} that for the paramagnetic solution of the IHM the presence of an intermediate metallic phase between the band insulator and Mott insulating phases is a real feature of the system.

Nevertheless, the reason that the intermediate metallic state is stabilized in the nonlocal CPA with respect to its local version when applied to strongly-correlated systems from an alloy-analogy point of view may be due to the fact that such theories of disorder fail to capture the localized state and hence always favour a metallic solution (see, for example, Ekuma~\etal\cite{Ekuma2013} and references therein).
Indeed, for $\Delta=0$ we would have expected the inclusion of intersite spatial correlations to actually lower the critical value $U_{c2}$ for the MIT in accordance with calculations for the paramagnetic solution of the conventional 2D Hubbard model using the variational cluster approximation~\cite{Balzer2009}, dynamical cluster approximation (DCA)~\cite{Moukouri1} and cellular-DMFT~\cite{Zhang2007,Park2008}, the latter two being the dynamical analogs of the nonlocal CPA and molecular CPA, respectively, for interacting electron systems. In the case of the DCA, it was found that in the weak coupling region a pseudogap at the Fermi level was systematically developed with increasing cluster size and its formation was attributed to the increasing length of intersite anti-ferromagnetic correlations~\cite{Moukouri1,Kyung2003}. Eventually, a true gap opens at zero temperature for any value of $U$ as the correlation length goes to infinity~\cite{Imada1998}. In the case of cellular-DMFT, it was reported that the MIT appears earlier than that obtained from DMFT due to the inclusion of short-range antiferromagnetic correlations~\cite{Zhang2007}. 

In conclusion, since our calculations are within the alloy analogy approach (i.e.~purely static),  we have been able to isolate the effects of nonlocal spatial correlations from the dynamics and found the opposite behaviour to that obtained with dynamical methods when nonlocal spatial correlations are included, i.e.~systematic enhancement of the metallic phase as opposed to systematic destruction. This important finding indicates that there must be some interplay between nonlocal spatial correlations and dynamical correlations which account for these opposing trends. This is a very interesting issue to investigate as future work.

%--------------------------------------------------------------------------------------------------------------------------------------------
\appendix
\section{Molecular CPA for the ionic Hubbard model}

After dividing the lattice into clusters or ``molecules'', the Green's function matrix in $\b{k}$-space for any given cluster can be expressed in the form
\begin{eqnarray}
	\bar{\u{\u{G}}}(\b{q},\omega) &=& \left[ \begin{array}{cc} \bar{\u{G}}_{AA}(\b{q}) &  \bar{\u{G}}_{AB}(\b{q}) \\  \bar{\u{G}}_{BA}(\b{q})  &  \bar{\u{G}}_{BB}(\b{q}) \end{array} \right] \nonumber\\
																												&=& \left[ \begin{array}{cc} \u{\omega}-(\u{\Sigma}_{AA}+\u{t}')-\u{t}'(\b{q}) & -(\u{\Sigma}_{AB}+\u{t})-\u{t}(\b{q}) \\ -(\u{\Sigma}_{BA}+\u{t})-\u{t}(\b{q}) &
																																	\u{\omega}-(\u{\Sigma}_{BB}+\u{t}')-\u{t}'(\b{q}) \end{array} \right]^{-1}
\end{eqnarray}
where for convenience the cluster sites have been labeled by grouping into sublattice components. Here $\b{q}$ has been used to denote the wavevector restricted to the BZ of the cluster superlattice and the intra-cluster hopping terms $\u{t}$ and $\u{t}'$ have been separated out from the definition of the cluster self-energy. The intra-sublattice hopping terms have been denoted by $\u{t}'$ since these are in general different from the inter-sublattice terms $\u{t}$. For nearest-neighbour hopping on the bipartite lattice, the intra-sublattice hopping terms are zero. The Green's function matrix can be expressed in real space by carrying out the BZ integration
\begin{equation}\label{mcpasc1}
	\bar{\u{\u{G}}}(\omega)=\frac{1}{\Omega_{BZ}}\int_{\Omega_{BZ}}\bar{\u{\u{G}}}(\b{q},\omega)\,d\b{q}
\end{equation}
where the integration is over the BZ of the cluster superlattice. Note that the inter-cluster hopping is expressed through the $\b{q}$-dependence of the hopping terms defined by
\begin{equation}
	\u{t}(\b{q})=\frac{N_c}{N}\sum_{C'\neq{C}}\u{t}\,e^{-i\b{q}(\b{R}_{C}-\b{R}_{C'})}
\end{equation}
where the distance vector implies the distance between the centres of clusters $C$ and $C'$, and for the case of nearest-neighbour hopping the appropriate terms in the matrix must be set to zero.

One may now proceed as in the nonlocal CPA by following the cavity matrix construction procedure expressed by equation (\ref{cavnlcpa}) through to (\ref{nlcpasc2}). Equation (\ref{mcpasc1}) and the analog of equation (\ref{nlcpasc2}) need to be solved self-consistently. It will be necessary to ensure that the resulting integrated DOS for each cluster site is consistent with the relevant site occupation number probability used in the cluster configurations of equation~(\ref{nlcpasc2}), thus adding an extra layer of self-consistency. In contrast to the nonlocal CPA, the real-space diagonal terms of $\u{\Sigma}_{\alpha\beta}$ and  $\u{G}_{\alpha\beta}$ when $\alpha=\beta$ do not have to be equivalent, and hence the DOS results will in general differ depending upon the position of the measured site even on the same sublattice.

For the cluster size $N_c=2$ the above formulation reduces to a generalisation of the coupled-CPA method of section \ref{sectioncpa}, albeit a different one to that of the nonlocal CPA due to the different boundary conditions imposed on the cluster.

%--------------------------------------------------------------------------------------------------------------------------------------------
\ack

It is an honour to produce this work in tribute to the late Prof.~Balazs Gy{\"{o}}rffy. This work is supported by National Natural Science Foundation of China (No.~11174219), Shanghai Pujiang Program (No.~11PJ1409900), Research Fund for the Doctoral Program of Higher Education of China (No.~20110072110044) and the Program for Professor of Special Appointment (Eastern Scholar) at Shanghai Institutions of Higher Learning as well as the Scientific Research Foundation for the Returned Overseas Chinese Scholars, State Education Ministry.

%--------------------------------------------------------------------------------------------------------------------------------------------
\section*{References}

\bibliographystyle{prsty}

%--------------------------------------------------------------------------------------------------------------------------------------------
\end{document}